\pdfoutput=1

\def\year{2018}\relax
\documentclass[letterpaper]{article} 
\usepackage{aaai18}  
\usepackage{times}  
\usepackage{helvet}  
\usepackage{courier}  
\usepackage{url}  
\usepackage{graphicx}  

\usepackage{amsmath}
\usepackage{subfig}
\usepackage{makecell}

\usepackage{natbib}
\usepackage[inline=true,margin=false]{fixme}
\usepackage{hyperref}

\DeclareMathOperator{\AAscore}{AA}
\DeclareMathOperator{\Katz}{Katz}
\DeclareMathOperator{\paths}{paths}

\nocopyright

\begin{document}
%
\title{Leveraging Friendship Networks for Dynamic Link Prediction in \\
Social Interaction Networks}
\author{Ruthwik R.~Junuthula, Kevin S.~Xu, and Vijay K.~Devabhaktuni\\
EECS Department, University of Toledo\\
2801 W. Bancroft St. MS 308, Toledo, OH 43606-3390, USA\\
\url{rjunuth@utoledo.edu}, 
\url{kevin.xu@utoledo.edu}, \url{vijay.devabhaktuni@utoledo.edu}}
\maketitle
\begin{abstract}
On-line social networks (OSNs) often contain many different types of 
relationships between users. 
When studying the structure of OSNs such as Facebook, two of the most 
commonly studied networks are friendship and interaction networks. 
The link prediction problem in friendship networks has been heavily studied. 
There has also been prior work on link prediction in 
interaction networks, independent of friendship networks. 
In this paper, we study the predictive power of \emph{combining friendship and 
interaction networks}. 
We hypothesize that, by leveraging friendship networks, we can improve the 
accuracy of link prediction in interaction networks.
We augment several interaction link prediction algorithms to incorporate
friendships and predicted friendships. 
From experiments on Facebook data, we find that incorporating 
friendships into interaction link prediction algorithms results in 
higher accuracy, but incorporating predicted friendships 
does not when compared to incorporating current friendships.
\end{abstract}

\section{Introduction}
Many different types of relationships between people are captured in 
online social networks (OSNs). 
For instance, Facebook captures friendships, wall posts, 
comments, likes, tags, and many other relations. 
Each type of relation can be used to construct a different type of 
network over the same set of nodes (people), with edges or links 
denoting the type of relation. 
These different types of networks can be categorized into two main types with 
distinct temporal dynamics:
\begin{itemize}
\item \emph{Friendship networks}, where edges denote some form of friendship, 
acquaintance, family relation, or perhaps simply an expression of interest 
in a person, i.e.~a follow.

\item \emph{Interaction networks}, where edges denote some form of 
interaction between nodes, such as having a conversation on a particular 
day.
\end{itemize}
In both cases, edges can be either directed or undirected depending on the 
type of friendship or interaction.

Dynamic friendship networks evolve slowly over time and are typically 
growing networks; that is, people add new friends much more often than 
they remove existing friends so that the networks densify over time 
\citep{Wilson2012}. 
On the other hand, dynamic interaction networks are highly variable over 
time. 
Two people may interact with each other at a certain time then cease to 
interact for a variety of reasons while still maintaining their friendship 
tie. 
On an OSN, an edge in an interaction network that persists over multiple 
time snapshots requires repeated interaction over time whereas an edge in a 
friendship network often requires only a one-time acknowledgment of a 
friendship or acquaintance. 

In this paper, we examine the problem of \emph{using friendship networks to 
improve predictions of future edges in interaction networks}. 
Since friendship networks are growing networks, predicting future edges in 
friendship networks requires only predicting the new edges that may appear 
in the future. 
On the other hand, interaction networks are evolving networks where nodes 
and edges are both added and deleted over time as interactions between 
people are initiated and dissolved, so predicting future edges in 
interaction networks requires 
predicting both the new edges that may appear as well as the current edges 
that may disappear.

We pose two main research questions in this paper. 
First, does incorporating the current friendship 
network lead to a more accurate prediction of the future interaction network? 
Second, does incorporating a predicted friendship network lead to a more 
accurate prediction of the future interaction network? 
We propose several methods of combining friendship and interaction networks 
to investigate these two questions on a Facebook data set 
\citep{Viswanath2009}. 

We find that either incorporating the current friendship network or the 
predicted friendship network leads to a more accurate prediction of the 
future interaction network compared to not using any friendship information at 
all. 
We observe this for 4 different interaction link predictors combined with 2 
different friendship link predictors. 
However, we find that incorporating predicted friendships does 
not improve link prediction accuracy for the interaction network compared to 
incorporating current friendships. 
This is due to the predicted friendships adding too many false positives that 
outweigh the added true positives they contribute. 

\section{Related Work}
There have been many studies on both the structures of friendship networks, 
also referred to as social graphs, and interaction 
networks, also referred to as activity networks, in OSNs. 
Past examinations of friendship networks have included measurements 
\citep{Mislove2007} and models of their growth \citep{Leskovec2008}, while 
past examinations of interaction networks have focused on persistence of 
interactions over time \citep{Viswanath2009}. 
There have also been examinations on the resemblance of the friendship and 
interaction networks on Facebook \citep{Wilson2012}.

\subsection{Link Prediction on Friendship Networks}
\label{sec:FriendLP}
Friendship networks are growing networks that densify over time 
\citep{Wilson2012} as many more friends are added than removed. 
Thus, the ``traditional'' link prediction setting where the objective is 
only to predict which new edges will form \citep{Liben-Nowell2007} is 
well-suited for predicting future friendships. 
The traditional link prediction problem has been extensively studied, and 
a variety of both supervised and unsupervised algorithms have been 
proposed; see \citet{Lu2011} for a survey of methods. 

We consider two simple yet 
effective unsupervised algorithms from the literature: Adamic-Adar \citep{adamic2003}, which we 
abbreviate as AA, and Katz. 
The link prediction scores of AA and Katz are calculated as follows \citep{Liben-Nowell2007}:
\begin{gather}
\label{eq:AA}
\AAscore(a,b) =  \sum_{c \in \Gamma(a) \cap \Gamma(b)} \frac{1}{\log \Gamma(c)} \\
\label{eq:Katz}
\Katz(a,b)    =  \sum_{l=1}^{\infty}\beta^l |\paths_{a,b}^{<l>}|
\end{gather}
where $\Gamma(c)$ denotes the neighbors of node $c$, $|\paths_{a,b}^{<l>}|$ 
denotes the number of paths of size $l$,
and $\beta \in (0,1)$ is a weight applied to lengths of paths. 

\subsection{Link Prediction on Interaction Networks}
\label{sec:InteractionLP}
Unlike friendship networks, edges are both \emph{added 
and removed} over time in interaction networks as people may interact for 
a period of time, cease to interact, and then resume their interactions.
Thus, the link prediction problem on interaction networks involves 
predicting both the new edges that will form and the existing edges that 
will persist. 
This more complex problem is often referred to as 
\emph{dynamic link prediction} \citep{xu2014dynamic}. 

The dynamic link prediction problem has also gained some recent attention, 
and most methods fall into one of three categories: univariate time series 
models, similarity-based methods, and probabilistic generative models 
\citep{Junuthula2016}.
We consider several dynamic link 
prediction algorithms that cover each of the three aforementioned 
categories: an exponentially-weighted moving average 
(EWMA), which is a special case of a general ARIMA univariate time series 
model \citep{Huang2009}; time series versions of Adamic-Adar 
(TS-AA) \citep{Gunes2015} and Katz (TS-Katz) \citep{Junuthula2016} that 
apply the EWMA to the AA and Katz scores in \eqref{eq:AA} and \eqref{eq:Katz}, 
respectively; and 
the dynamic stochastic block model (DSBM), a probabilistic generative model, 
combined with the EWMA \citep{xu2014dynamic}. 
The EWMA alone is simply a summary of past interactions and thus does not 
predict any new interactions, unlike the other three methods.

\subsection{Link Prediction on Multiple Networks}
More recent work on link prediction has involved the use of multiple 
networks or other data sources. 
\citet{dong2015} and \citet{hristova2016} proposed methods for link prediction 
across multiple networks, where they predict missing links in one network using 
links the other network. 
\citet{Gong2014} considered jointly inferring links between users and 
user attributes on Google+. 
\citet{Merritt2013} examined the problem of predicting friendships between 
players of a multi-player video game using their interactions, such as 
playing on the same team. 
None of this work considers the fundamental differences in 
structure and temporal dynamics between friendship and interaction networks
(growth vs.~evolution), 
which we consider in this paper to predict future interactions using 
both interactions and friendships. 

\section{Data Description}
We investigate the data set collected by \citet{Viswanath2009} by 
crawling the Facebook New Orleans regional network. 
The data set contains friendships and wall posts of over $60,000$ users from 
September 2006 to January 2009, along with timestamps for all wall posts and 
for friends added after the initial crawl. 
We construct a sequence of network snapshots over $90$-day intervals, 
similar to several other analyses of the data 
\citep{Viswanath2009,Junuthula2016}, 
from the start of the data trace to the last 
full snapshot that ends in November 2008, resulting in a total of $9$ 
snapshots.
At each snapshot, we create two adjacency matrices, one with edges 
representing friendships, and one with edges representing interactions (wall 
posts) occurring during the snapshot. 

\begin{figure}[t]
\centering
\subfloat[]{
\includegraphics[width=0.45\columnwidth]{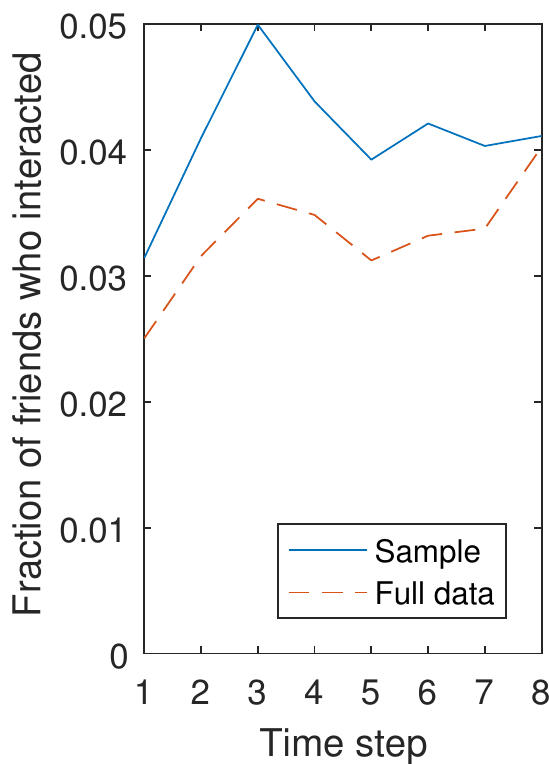}
\label{fig:Friends_that_Interacted}
}
\quad
\subfloat[]{
\includegraphics[width=0.45\columnwidth]{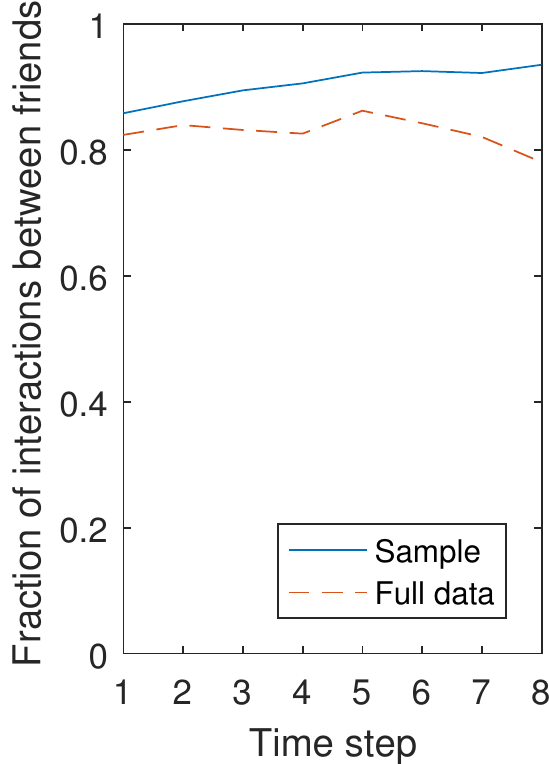}
\label{fig:Interactions_btw_friends}
}
\caption[]{\subref{fig:Friends_that_Interacted} Fraction of friends at time $t$ who interacted at time $t+1$.
\subref{fig:Interactions_btw_friends} Fraction of interactions at time $t+1$ that are between friends at time $t$. 
In both cases, the fraction computed on the sample we study ($1,911$ nodes) are only slightly higher than on the full data ($> 60,000$ nodes).}
\label{fig:FriendsInteractions}
\end{figure}

In this paper, we study a representative sample of the full data set consisting 
of all users with degree $120$ or higher in the aggregated friendship network 
over the entire data trace, resulting in networks containing $1,988$ nodes. 
Figure \ref{fig:Friends_that_Interacted} shows that only a small 
fraction of friends interact via wall posts during any given $90$-day 
time snapshot. 
This is partially due to 
the friendship network being much denser than the interaction network. 
Figure \ref{fig:Interactions_btw_friends} shows that the overwhelming 
majority of interactions between users occur between friends. 
Both fractions are only slightly higher in the sample we analyze compared to 
the full data set. 

\section{Methodology}

Given the relationship between friendship and interaction networks seen in 
Figure \ref{fig:FriendsInteractions}, one might expect friendship networks to be 
useful for link prediction on interaction networks. 
Thus, we pose the following research questions:
\begin{description}
\item[RQ1] Does incorporating the friendship network at the current time 
snapshot $t$ lead to a more accurate prediction of the future interaction 
network (at time $t+1$) compared to just using the interaction 
networks up to time $t$?

\item[RQ2] Does incorporating also the predicted friendship network at time 
$t+1$ (along with the current friendship network) lead to a more accurate 
prediction of the future interaction network (at time $t+1$)?
\end{description}

Our objective is \emph{not} to develop a new 
link prediction algorithm for friendship networks or for interaction 
networks. 
Rather, we are interested in how we can incorporate information from 
current or predicted friendship networks into predicting future interaction 
networks, which would allow us to answer our two research questions.

\subsection{Using Friendships to Predict Interactions}
We consider two approaches to incorporate the friendship network into dynamic 
link predictions on the interaction network. 
The first approach is to use \emph{current friendships} (up to time snapshot 
$t$) to 
inform the prediction of future interactions (at time $t+1$). 
We do this by taking a convex combination of the adjacency 
matrix of the friendship network at time $t$ and the matrix of link 
prediction scores obtained from the interactions (using a dynamic link 
prediction algorithm as described in the \nameref{sec:InteractionLP} 
section) up to time $t$\footnote{The scores from the AA and Katz 
link predictors are normalized to the 
interval $(0,1)$ before taking the convex combination to put them on the 
same scale as the friendship adjacency matrix.}. 
This convex combination is then used as the matrix of link prediction scores 
for future interactions at time $t+1$. 
These scores are then compared to the interactions that actually take place 
at time $t+1$ (see the \nameref{sec:Eval} section for details). 

The second approach we consider is to use \emph{predicted friendships} to 
inform the prediction of future interactions. 
We do this by running a traditional link prediction algorithm (described in 
the \nameref{sec:FriendLP} section) on the friendship network at time $t$ and 
replacing the zeros in the friendship adjacency matrix with the (normalized) 
friendship link prediction scores. 
We then take a convex combination of this matrix with the matrix of dynamic 
link prediction scores from the interactions.

\subsection{Evaluation Metrics}
\label{sec:Eval}
Evaluating link prediction accuracy involves comparing a binary 
label (whether or not an edge occurs) with a real-valued predicted 
score from the link prediction algorithm.  
Evaluation in such a setting typically involves computing the area under 
a threshold curve such as 
the area under the Receiver Operating Characteristic (ROC) curve, typically 
referred to just as AUC, or the area under the Precision-Recall (PR) curve, 
which we denote by PRAUC. 

\citet{Junuthula2016} studied evaluation metrics for the dynamic link 
prediction problem and recommended splitting the evaluation into new link 
prediction using PRAUC and previously observed link prediction using AUC, due 
to the massive difference in the degree of difficulty of the two problems. 
These two quantities were then combined into a single balanced metric 
using the geometric mean (following a correction for chance) denoted by 
the GMAUC. 
We adopt these three metrics for evaluating predictions of future 
interactions. 

\section{Results and Discussion}

The accuracy metrics for predicting future interaction networks are shown 
in Table \ref{tab:AccInteractions}. 
The interaction network link predictors are split into three categories: 
predictors that do not use friendships, predictors that use only current 
friendships (appended with FR), and predictors that use predicted 
friendships (appended with the friendship link predictor used). 
For predictors that use friendships, we perform a grid search over all 
possible convex combinations in increments of $0.1$ and report the results 
with the highest GMAUC\footnote{Code and data to reproduce our experiment 
results are available at 
\url{https://github.com/IdeasLabUT/Friendship-Interaction-Prediction}.}.

\begin{table}[t]
	\renewcommand{\arraystretch}{1.1}
	\centering
	\caption{Accuracy metrics for prediction of interaction links separated 
	into new and previously observed links along with combined GMAUC metric. 
    The first four methods use no friendships, the next four use current 
    friendships (+ FR in the name), and the last eight use 
    predicted friendships (predictor indicated by + AA or + Katz in 
    the name).}
    \label{tab:AccInteractions}
	\begin{tabular}{cccc}
		\hline
		Predictor      & \makecell{PRAUC \\ (new)} & \makecell{AUC \\ (previous)} & GMAUC \\
		\hline
		EWMA             & 0.001         & 0.698         & 0.000 \\
		TS-AA            & 0.011         & 0.577         & 0.040 \\
		TS-Katz          & 0.012         & 0.600         & 0.046 \\
		DSBM             & 0.004         & 0.701         & 0.032 \\
		\hline
		EWMA + FR        & 0.028         & 0.699         & 0.104 \\
		TS-AA + FR       & 0.043         & 0.574         & 0.079 \\
		TS-Katz + FR     & 0.051         & 0.597         & 0.098 \\
		DSBM + FR        & 0.037         & 0.701         & 0.121 \\
		\hline
		EWMA + AA        & 0.026         & 0.696         & 0.100 \\
		EWMA + Katz      & 0.027         & 0.696         & 0.100 \\
		TS-AA + AA       & 0.037         & 0.574         & 0.073 \\
		TS-AA + Katz     & 0.037         & 0.575         & 0.073 \\
		TS-Katz + AA     & 0.037         & 0.569         & 0.071 \\
		TS-Katz + Katz   & 0.037         & 0.569         & 0.071 \\
		DSBM + AA        & 0.026         & 0.696         & 0.100 \\
		DSBM + Katz      & 0.027         & 0.696         & 0.101 \\
		\hline
	\end{tabular}
\end{table}

We begin by investigating RQ1: whether incorporating current friendships 
improves predictions of interactions. 
For each of the four interaction link predictors, we see a substantial 
improvement in overall link prediction accuracy, as indicated by the GMAUC, 
by incorporating current friendships compared to no friendships. 
By examining the PRAUC (new) and AUC (previously observed) values, 
we see that this is 
accomplished primarily by increase accuracy in predicting new edges, 
sometimes by trading off accuracy in predicting previously observed 
edges.
This is a reasonable result because the inclusion of friendships should 
increase the number of false positives while also leading to 
a significantly greater number of correctly predicted new edges. 
Thus, it certainly appears that the answer to RQ1 is 
yes---\emph{incorporating current friendships does indeed result in a 
significantly better link predictor for interaction networks}. 

Next, we investigate RQ2: whether incorporating predicted friendships improves 
predictions of interactions. 
For each of the four interaction link predictors, we do see an improvement in 
accuracy compared to using no friendships; however, we actually see a 
\emph{decrease} 
in accuracy, both in predicting new and previously observed edges, 
compared to using current friendships, 
regardless of whether AA or Katz is used as the friendship link predictor. 
We find this to be due to predicted friendships adding an overwhelming amount of false positives compared to the added true positives, resulting in lower GMAUC compared to adding only current friendships.
Thus, the answer to RQ2 appears to be no---\emph{incorporating predicted 
friendships 
offers an improvement compared to no friendships but not compared to using 
current friendships}. 

For almost all of the predictors, we find that assigning a weight of $0.9$ to interaction predictors and $0.1$ to friendships (current or predicted) results in the highest GMAUC. 
Hence, the structures of current and past interaction networks appear 
to produce a much better 
predictor of future interactions than the structures of friendship networks; 
however, 
our results demonstrate that incorporating friendship networks can definitely 
improve accuracy of interaction link predictions. 
Another interesting observation is that the EWMA and DSBM benefit more 
from the addition of friendships than TS-AA and TS-Katz. 
We believe that this is due to the EWMA and DSBM lacking mechanisms for 
predicting triadic closure, which is a key element of link prediction 
(for both interactions and friendships) in OSNs. 
Adding the friendship network, which includes many triangles, provides them  
some mechanism to favor triadic closure, which greatly improves the 
accuracy for both predictors.

\bibliographystyle{aaai}
\bibliography{ICWSM_2018_LinkPrediction}

\end{document}